\documentclass[twoside]{dis07}
\usepackage[latin1]{inputenc}
\usepackage[dvips]{graphicx,epsfig,color}
\usepackage{wrapfig,rotating}
\usepackage{amssymb,amsmath,array}
\usepackage{cite}

\newcommand{\gsim}{\raisebox{-0.07cm}{$\:\:\stackrel{>}{{\scriptstyle
 \sim}}\:\: $} }
\newcommand{\lsim}{\raisebox{-0.07cm}{$\:\:\stackrel{<}{{\scriptstyle
 \sim}}\:\: $} }
\newcommand{\beq}{\begin{equation}}
\newcommand{\eeq}{\end{equation}}
\newcommand{\bea}{\begin{eqnarray}}
\newcommand{\eea}{\end{eqnarray}}
\newcommand{\nn}{\nonumber}
\newcommand{\MSb}{$\overline{\mbox{MS}}$}
\newcommand{\as}{\alpha_{\rm s}}

\def\nf{{n^{}_{\! f}}}

\begin{document}
\vspace{-5mm}
\title{
{\vspace*{-1.4cm}\small LTH 752 \hfill 
{\tt arXiv:0707.4106 [hep-ph]}}\\[1.5cm]
{\large Parton Distributions: Progress and Challenges}\\[-2mm]}

\author{Andreas Vogt
\vspace{.15cm}\\
Department of Mathematical Sciences, University of Liverpool \\
Liverpool L69 3BX, United Kingdom \\[-1.5mm]
}

\maketitle

\begin{abstract}
We briefly discuss recent research on the spin-averaged parton densities of the
proton, focusing on some aspects relevant to hard processes at the LHC. 
Specifically, after recalling the basic framework and the need for higher-order
calculations, we address the evolution equations governing the scale dependence
of the parton distributions and their solution, schemes for initial conditions
and the inclusion of heavy quarks, recent progress on fits to data, and future
high-precision constraints from LHC measurements.\\[-4mm] 
\end{abstract}

\section{Introduction: partons for the LHC}

\vspace*{-1mm}
For at least the next ten years, proton$\,$--(anti-)$\,$proton colliders will 
continue to form the high-energy frontier in particle physics. At such 
machines, many quantitative studies of hard (high mass/scale) standard-model 
and new-physics processes require a precise understanding of the parton 
structure of the proton. The present talk \cite{url} briefly discusses some 
recent developments in this field. 

\begin{wrapfigure}{r}{0.45\columnwidth}
\vspace*{-0.4cm}
\centerline{\includegraphics[bb = 170 560 430 360, scale = 0.6]
{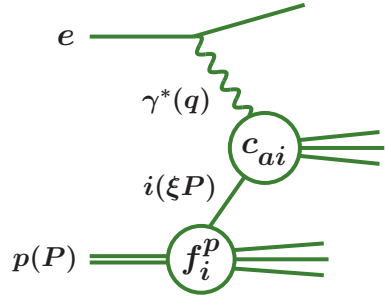}}
\vspace*{4cm}
 \caption{Kinematics of photon-exchange DIS in the QCD-improved parton model. 
 Particle momenta are given in brackets.}\label{AVfig1}
\vspace*{-0.3cm}
\end{wrapfigure}

\vspace{0.5mm}
We start by recalling the description of hard proton processes using the 
simplest case, inclusive lepton-proton deep-inelastic scattering (DIS), the 
process providing the major part of the present constraints on the parton 
densities. Here the hard scale is the virtuality $\, Q^2 = - q^2 \,$ of the 
exchanged gauge boson, a photon in Fig.~\ref{AVfig1}, and the Bjorken variable 
$\, x = Q^2 / (2 P q)\,$, with $P$ the proton momentum, is usually 
chosen as the second independent variable. At zeroth order in the strong
coupling constant $\alpha_{\rm s}$ the hard coefficient functions $c_{a,i}$ are 
trivial, and the momentum fraction $\xi$ carried by the struck quark $i$ is
equal to Bjorken-$x$ if mass effects are neglected.

\vspace{1mm}
In general, the structure functions $F^{\,p}_{2,L}$ for the process of 
Fig.~\ref{AVfig1} are given by 
 
\vspace{-2.5mm}
\bea
\label{AVeq1}
  x^{\,-1} F_a^{\,p}(x,Q^2) \; = \; \sum_{i\, =\, q, g} \:\int_x^1 
  \! \frac{d\xi}{\xi} \:  c^{}_{a,i}\bigg( \frac{x}{\xi}, \as(\mu^2), 
  \frac{\mu^2}{Q^2} \bigg) \: f_{i}^p(\xi,\mu^2) 
\eea
 
\vspace{-0.5mm}
\noindent
plus terms of order $1/Q^2$ which, for the purpose of high-scale predictions,
are best suppressed by sufficiently stringent cuts on the fitted experimental
data. Besides on the factorization scheme used to define the parton densities
$f_{i}^p$ -- in this talk \MSb\ unless stated otherwise -- the coefficient
functions depend on the renormalization and factorization scale $\mu$ 
(identified here for notational simplicity) which ought to be chosen as
$\mu^2 = {\cal O}(Q^2)$ in order to avoid large logarithms. The parton
distributions depend on this scale via the evolution equations
\beq
\label{AVeq2}
  \frac{d}{d \ln \mu^2} \, f_i^{}(\xi,\mu^2) \; =\;
  \sum_{k} \left[ P^{}_{ik}(\as(\mu^2)) \otimes f_k^{}(\mu^2) \right](\xi)
  \:\: .
\eeq
Here $\otimes $ is a short-hand for the Mellin convolution written out in 
Eq.~(\ref{AVeq1}) above. The initial conditions for Eq.~(\ref{AVeq2}) are, of
course, not calculable in perturbative QCD. As lattice results are restricted
to very few Mellin moments (with, at present, still rather limited accuracy), 
predictions for collider cross sections are obtained via fits to suitable sets
of reference observables, including structure functions in DIS, and the 
universality of the parton densities.

\vspace{1mm}
The splitting functions $P$ and the process-dependent hard coefficient
functions $c_a$ admit expansions in powers of $\as$,
\bea
\label{AVeq3}
 P \, &\! =\! &
    \as\, P^{(0)} \, +\: \as^2\, P^{(1)}
    \, +\: \as^3\, P^{(2)} +\, \ldots \quad   \nn \\[1mm]
 c^{}_{a} &\! =\! & \as^{\,n_a}
   \big[ \, c_a^{(0)} \, + \, \as\, c_a^{(1)}
   + \: \as^2\, c_a^{(2)} \, +\, \ldots \,\big] 
\eea
with, for example, $n_a = 0$ for $F_2$ and $n_a = 1$ for $F_L$.
For a consistent approximation the same number of terms has to be kept in the
two lines of Eq.~(\ref{AVeq3}). The first $n\! +\! 1$ terms define the 
N$^{\rm n}$LO approximation. As the normalization of the LO prediction is
rather arbitrary, the next-to-leading order (NLO) provides the first real 
prediction of the cross sections and, consequently, the NNLO the first serious 
error estimate of the perturbative expansions. 
 
\begin{figure}[hb]
\vspace{1mm}
\centerline{\epsfig{figure=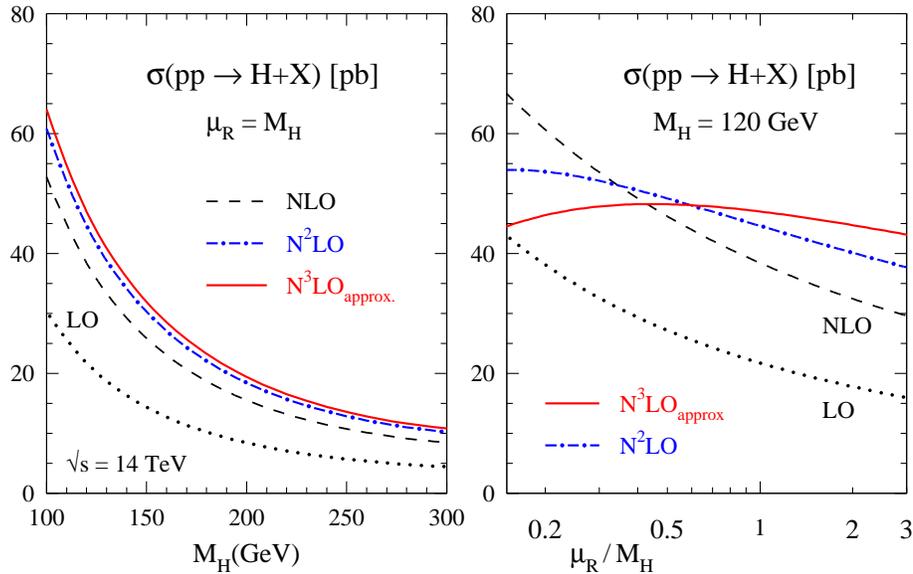,width=12.5cm}\hspace*{3mm}}
\vspace{-1mm}
\caption{Perturbative expansion of the total cross section for Higgs production
 at the LHC. Shown are the dependence on the mass $M_H$ and the 
 renormalization scale $\mu_{\rm r}$ (from Ref.~\cite{Moch:2005ky}).}
 \label{AVfig2}
\vspace*{-4mm}
\end{figure}

\vspace{1mm}
The successive approximations of perturbative QCD are illustrated in 
Fig.~\ref{AVfig2} for an LHC process of utmost importance, the production of 
the standard-model Higgs boson dominated by gluon-gluon fusion via a top-quark
loop. Obviously the NLO approximation \cite{HiggsNLO1} is insufficient for
a quantitative prediction in this case, and even at NNLO 
\cite{HiggsNNLO1,Anastasiou:2005qj} 
higher-order uncertainties of about 15\% remain for the total cross section. 
A perturbative accuracy of 5\% is only reached at N$^3$LO, known to a 
sufficient approximation from Ref.~\cite{Moch:2005ky} (for an extension to 
the rapidity distribution see Ref.~\cite{Ravindran:2006bu}). Note that these
uncertainties do not include those of the coupling $\as$ and the parton 
densities, taken for Fig.~\ref{AVfig2} from Ref.~\cite{MRST0102} where at NNLO
previous (but sufficiently accurate) approximations~\cite{vanNeerven:2000wp} 
were used for $P^{(2)}$ in Eq.~(\ref{AVeq3}).

\begin{wrapfigure}{r}{0.55\columnwidth}
\vspace*{-0.5mm}
\hspace*{-2mm}\centerline{\epsfig{file=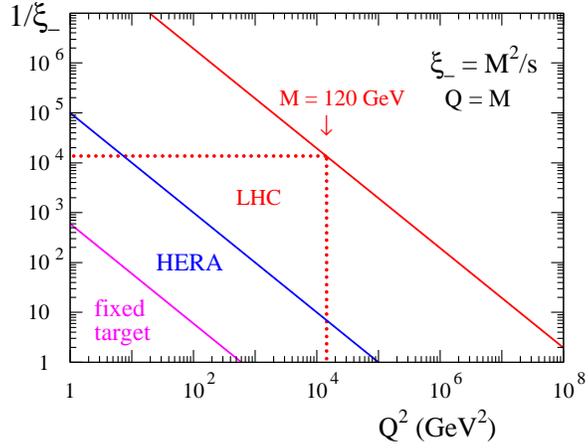,width=8.2cm}}
\vspace*{-2.5mm}
 \caption{Minimal parton momenta $\xi_{-}$ probed at the LHC, compared with 
 the DIS coverage of HERA and previous fixed-target experiments.}\label{AVfig3}
\vspace*{-3mm}
\end{wrapfigure}

The minimal momentum fractions $\xi^{}_{-}$ of partons contributing to the 
production of a particle of mass $M$ at the LHC are shown in Fig.~\ref{AVfig3},
together with the kinematic reach of HERA and fixed-target DIS experiments 
at the corresponding scales $Q^2$. 
Taking into account also the limited rapidity coverage of the LHC detectors, 
one reads off $\xi^{}_{-} \!\gsim 10^{-4}$ for the most important processes,
including the production of the $W\!$, $Z$ and Higgs bosons and the top quark, 
and the search for new particles. Thus the HERA data can be fitted with a cut 
of $\,Q^2 \approx 10 \mbox{ GeV}^{\:\!2}\,$ which should be sufficient to 
suppress low-scale instabilities (as, e.g., in $F_L$ to NNLO 
\cite{Moch:2004xu}) and power corrections to Eq.~(\ref{AVeq1}). 

\vspace*{1mm}
\section{Higher orders in the parton evolution}

\begin{figure}[hb]
\vspace{1mm}
\centerline{\epsfig{figure=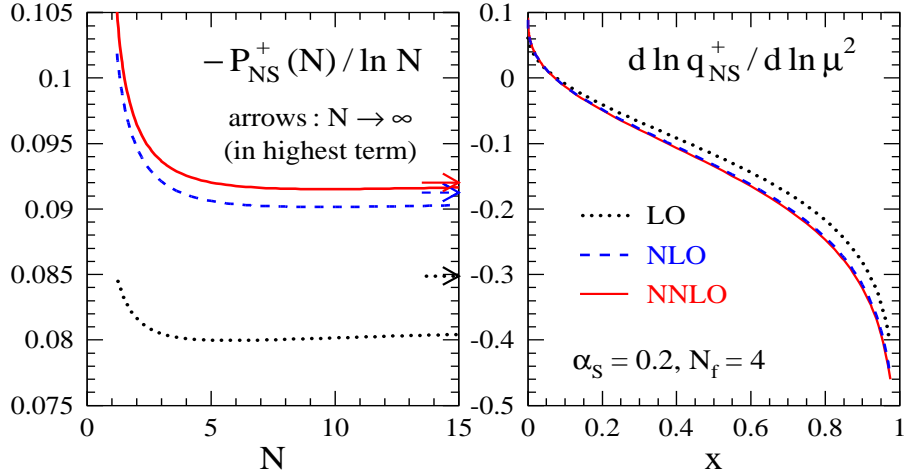,width=12cm,height=6.2cm}%
\hspace*{3mm}}
\vspace{-1mm}
\caption{The LO, NLO and NNLO  approximations to the splitting-function moments 
 $P_{\rm ns}^{\,+}(N)$ for four flavours at $\as = 0.2\,$, and the resulting 
 logarithmic scale derivatives for $xq_{\rm ns}^{\,+}\,=\, x^{\,0.5} (1-x)^3$,
 a schematic but characteristic model distribution (from Ref.~\cite
 {Moch:2004pa}).}\label{AVfig4}
\vspace*{-1mm}
\end{figure}

The complete NNLO splitting functions $P^{(2)}(x)$ -- from now on we, as usual,
denote also the parton momentum fractions by $x$ -- in Eq.~(\ref{AVeq3}) have 
been computed three years ago in Refs.~\cite{Moch:2004pa,Vogt:2004mw}. We first
consider the flavour non-singlet evolution of quark-distribution differences
such as the combination $\, q_{\;\!\rm ns}^{\;\! +} = u\!+\!\bar{u} - 
(d\!+\!\bar{d}\,)\,$ probed by $F_2^{\;\! p} - F_2^{\;\! n}$. 
Figure~\ref{AVfig4} illustrates the perturbative expansion of the Mellin 
moments of the corresponding splitting function and the resulting 
approximations for the scale dependence of $q_{\;\!\rm ns\,}^{\;\! +}$ at 
large $x$. 

The first fourth-order result for this splitting function has been presented 
last year \cite{Baikov:2006ai}: the second moment of $P_{\rm ns}^{\,+}(x)$ is 
now known to N$^3$LO for three flavours, numerically reading
\beq
\label{AVeq4}
  P_{\sf ns}^{\:\! +}(N\! = \! 2, \nf\! =\! 3) \:\: =\:\: 
  -\, 0.283\,\as\: [\, 1\, +\, 0.869\,\as\,
  +\, 0.798\,\as^2\, +\, 0.926\,\as^3 \, +\, \ldots\, ] \:\: .
\eeq
Taking into account the weak $N$-dependence of $P_{\rm ns}^{\,+}$ at $N\!>\! 2$
demonstrated in Fig.~\ref{AVfig4}, this result sets the scale for the N$^3$LO 
contributions for the whole large-$x$ region. According to the general pattern,
the corresponding corrections for $\nf = \, 4\,\ldots 6$ will be even smaller. 

\begin{figure}[hb]
\vspace*{-0.5mm}
\centerline{\epsfig{figure=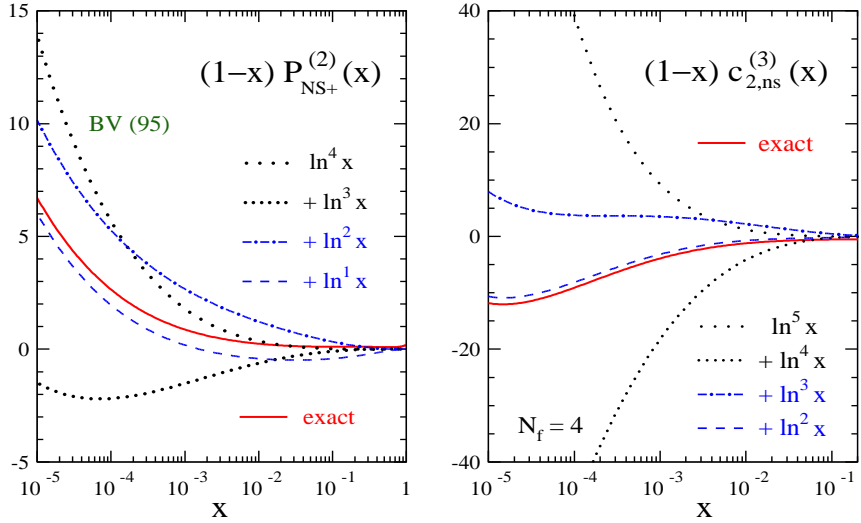,width=12cm,height=7.0cm}%
\hspace*{3mm}}
\vspace{-1.5mm}
\caption{The exact $\as^3$ contributions to the non-singlet splitting function
 and coefficient function for $F_2$, compared to approximations obtained from 
 the small-$x$ logarithms (from Refs.~\cite{Moch:2004pa,Vermaseren:2005qc}). 
 The leading small-$x$ term of $P_{\rm ns +}^{(2)}$ was derived before in 
 Ref.~\cite{Blumlein:1995jp}.}\label{AVfig5}
\vspace{-1mm}
\end{figure}

\vspace{1mm}
The low-$x$ behaviour of the non-singlet splitting functions and coefficient
functions is not too relevant in practice, but provides an interesting 
lab for the study of small-$x$ logarithms: unlike in the singlet case,
two additional powers of $\,\ln x\,$ enter per order in $\as$, e.g., terms up 
to $\ln^4 x$ and $\ln^5 x$ occur in $P_{\rm ns}$ and $c^{}_{\:\! 2,\rm ns}$ 
already at order $\as^3$. Successive approximations of these 
functions including the leading, next-to-leading, \dots small-$x$ terms are 
shown in Fig.~\ref{AVfig5}.
 
\noindent
Obviously, a `low-order' approximation of this type is not appropriate at any 
$x$-values relevant to colliders. Consequently leading- and next-to-leading-log 
resummations can, at best, provide only very rough indications of the maximal
size of the higher-order corrections. In the present case the all-order 
leading-logarithmic contributions \cite{Blumlein:1995jp} are small enough to 
exclude small-$x$ instabilities, e.g., for $\,xq_{\rm ns}^{\,+}\sim x^{\,0.5}$, 
down to extremely low values of $x$ \cite{Blumlein:1996dd}. 

\begin{figure}[thp]
\vspace*{0.5mm}
\centerline{\epsfig{figure=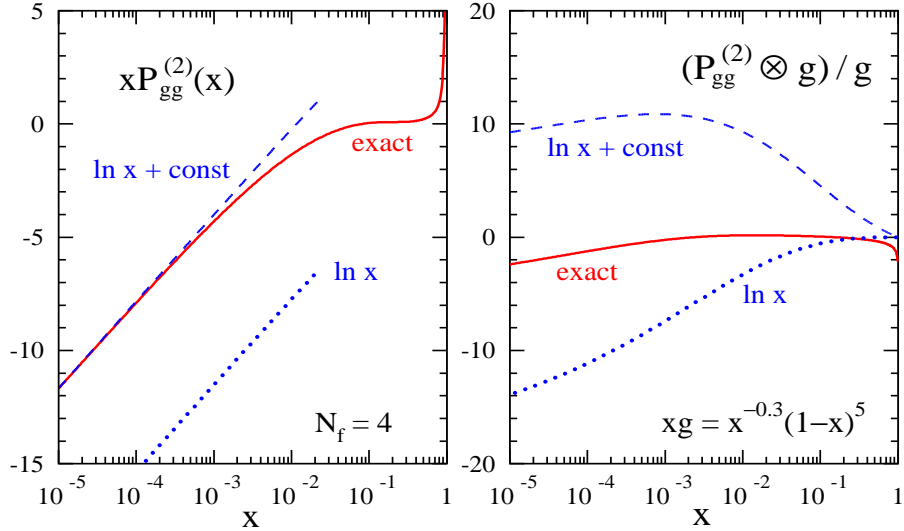,width=12cm,height=7.0cm}%
\hspace*{3mm}}
\vspace{-1.5mm}
\caption{Left: the $\as^3$ contribution $P_{\rm gg}^{(2)}$ to the gluon-gluon 
 splitting function, compared to its leading \cite{NL-BFKL,vanNeerven:2000uj} 
 and next-to-leading small-$x$ approximations. 
 Right: the convolution of these three functions with a schematic but typical 
 gluon distribution (from Ref.~\cite{Vogt:2004mw}).}\label{AVfig6}
\vspace*{-3mm}
\end{figure}

The flavour-singlet splitting functions are vital for transferring small-$x$
information from HERA to LHC scales across up to three orders of magnitude in
$Q^2$, recall Fig.~\ref{AVfig3}. The corresponding NNLO contribution 
\cite{Vogt:2004mw} to Eq.~(\ref{AVeq2}) is shown in Fig.~\ref{AVfig6} for the 
gluon-gluon~case.
Also here the leading small-$x$ term, obtained before in Ref.~\cite{NL-BFKL}
(and transformed to \MSb\ in Ref.~\cite{vanNeerven:2000uj}), does not provide
a good approximation for any practically relevant values of $x$. 
Moreover, the splitting functions enter physical quantities only via the Mellin 
convolutions of Eqs.~(\ref{AVeq1}) and (\ref{AVeq2}). Hence a locally accurate 
low-$x$ approximation, as provided for $P_{\rm gg}^{(2)}$ by the $\, x^{-1} 
\ln x\,$ plus the $\,x^{-1}$ terms at $\,x \lsim 10^{-3}$, is insufficient for
$dg/d\ln \mu^2$ even at small $x$. 

\begin{figure}[hbt]
\vspace{-4cm}
\centerline{\epsfig{figure=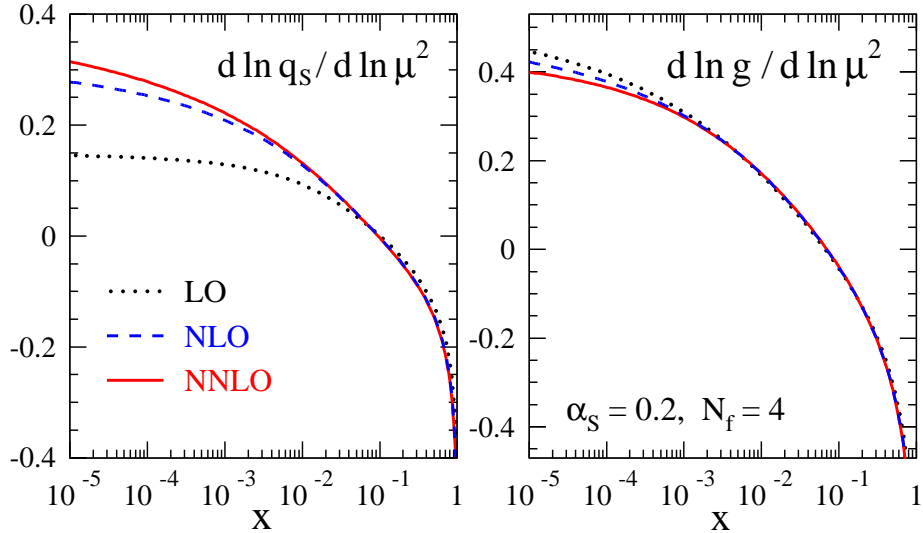,width=12.5cm,height=7.0cm}%
\hspace*{3mm}}
\vspace{-1.5mm}
\caption{Perturbative expansion of the scale derivatives of typical quark
 and gluon distributions at $\mu^2 \approx 30$ GeV$^2$ (from 
 Ref.~\cite{Vogt:2004mw}, where the initial conditions are specified).}
 \label{AVfig7}
\vspace*{-1mm}
\end{figure}

Consequently, reliable estimates of the post-NNLO corrections to the 
small-$x$ evolution will become possible, via approximations analogous to those
of  Ref.~\cite{vanNeerven:2000uj}, only once a few singlet moments have been 
computed to order $\as^4$. 
Fortunately, as illustrated in Fig.~\ref{AVfig7}, the expansion of the quark
and gluon evolution to NNLO appears to be very stable, at least for the main 
HERA-to-LHC region $x \gsim 10^{-4}$ at $\,Q^2 \gsim 10 \mbox{ GeV}^{\:\!2}$
(see above). 

\vspace{4.1cm}

\section{Solutions of the evolution equations}

\begin{figure}[hbt]
\vspace{-1mm}
\centerline{\hspace*{6mm}\epsfig{figure=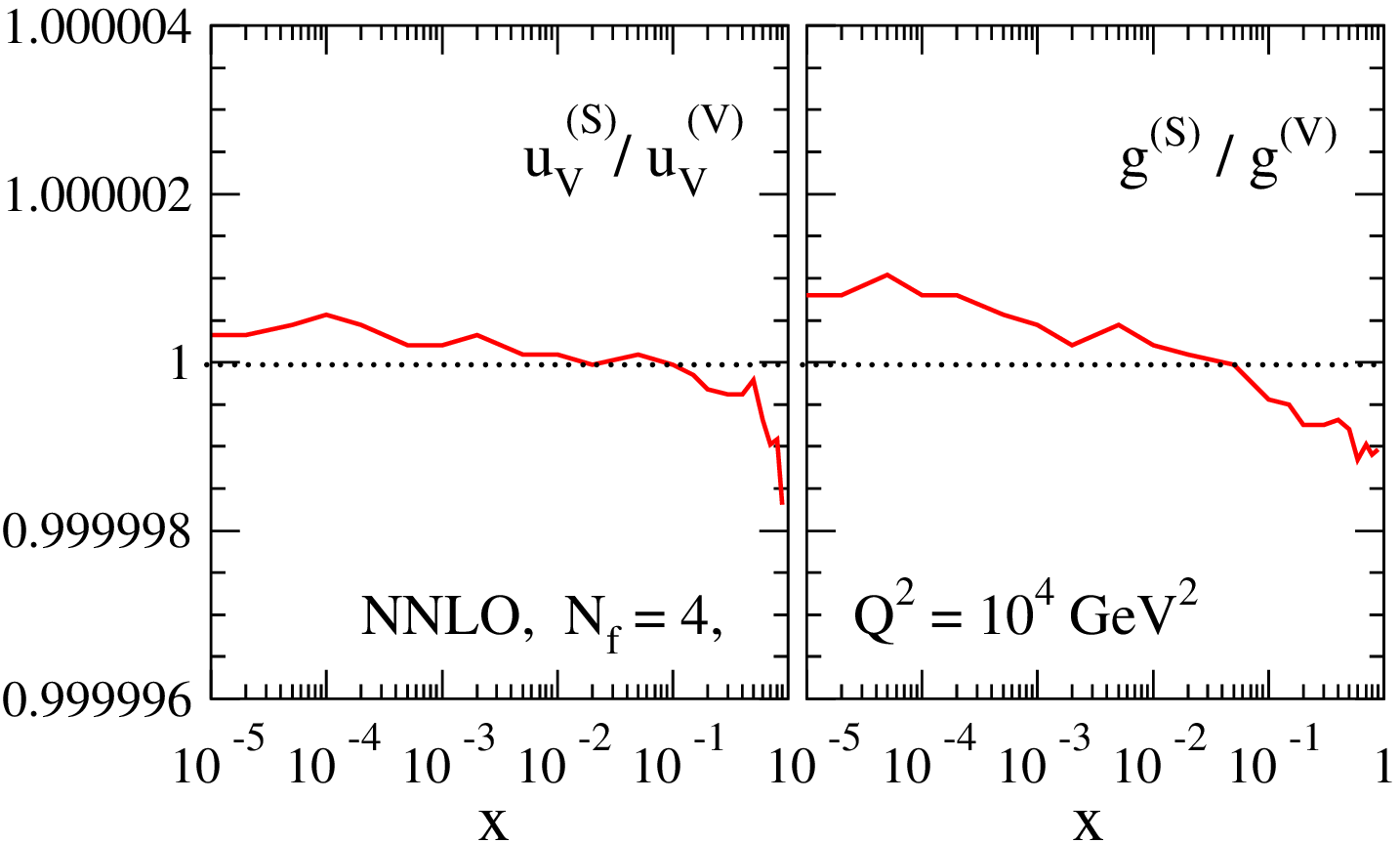,width=7.5cm,%
height=4.8cm} \quad \epsfig{figure=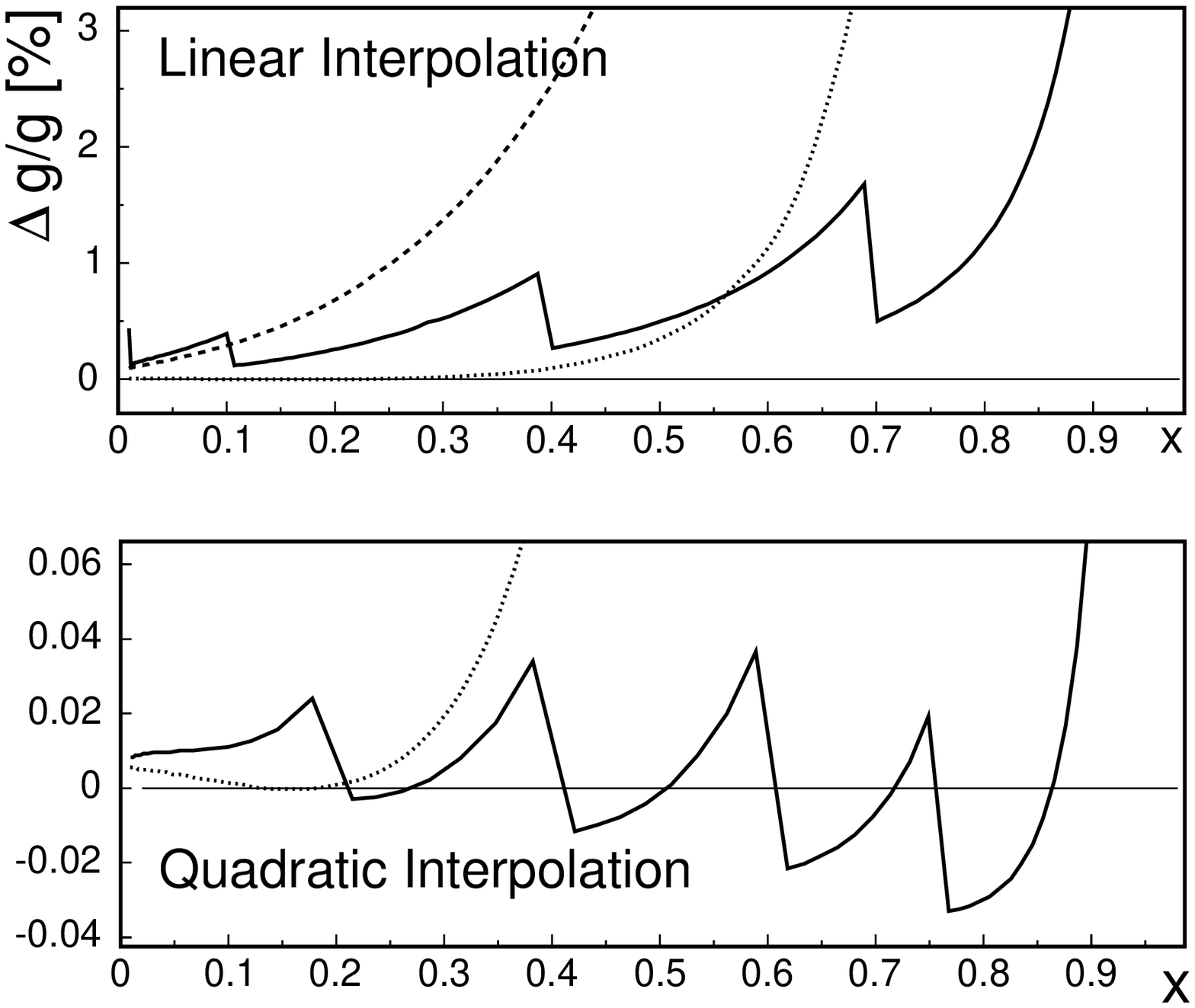,width=6.5cm}}
\vspace{-1mm}
\caption{Left: ratios of high-scale NNLO up-valence and gluon distributions 
 after evolution with the codes \cite{HOPPET} (S) and \cite{Pegasus} (V). 
 Right: the accuracy improvement of QCDNUM due to the new quadratic 
 $x$-interpolation with actually fewer points (for details see 
 Ref.~\cite{QCDNUM}).}
\label{AVfig8}
\vspace*{-0.5mm}
\end{figure}

The most direct manner to solve the system (\ref{AVeq2}) of coupled integro-%
differential equations is by a discretization in both $x$ and $\mu^2$. Recently
written or updated public codes including the NNLO evolution are HOPPET 
\cite{HOPPET} and the new version 17 (beta-released at the time of this talk) 
of QCDNUM \cite{QCDNUM}. 
Alternatively, Eqs.~(\ref{AVeq2}) can be transformed to ordinary differential 
equations in (complex) Mellin-$N$ space. These are then treated analytically 
and the solutions transformed back by quadratures. This approach has been 
employed in QCD-{\sc Pegasus} \cite{Pegasus}.
The left part of Fig.~\ref{AVfig8} shows a sample comparison of the programs 
\cite{HOPPET} and \cite{Pegasus}, using the Les Houches initial conditions 
discussed below. The right part of the figure, taken from the QCDNUM manual, 
illustrates the greatly improved numerical accuracy of the new version.  

Obviously it is very useful, e.g., for validating newly written or ported 
codes, to have at one's disposal a set of benchmark evolution results.
A reference input was set up for this at the 2001 Les Houches collider-physics 
workshop (see Ref.~\cite{PDF-ref} for the complete expressions),
\beq
\label{AVeq5}
  xu_{\rm v} (x,\mu_{\rm f,0}^2) \: = \: 5.1072\:\: x^{0.8}\:\: (1-x)^3
 \: , \: \ldots , \:
  xg (x,\mu_{\rm f,0}^2)  \: = \: 1.7000\, x^{-0.1}\, (1-x)^5 \quad
\eeq
for the initial factorization scale $\,\mu_{\rm f,0}^2 \, = \, 2$ GeV$^2\,$ 
and the coupling $\as (\mu_{\rm r}^2  = 2\mbox{ GeV}^2) \, = \, 0.35\, $.
As illustrated in Fig.~\ref{AVfig8}, the results of the programs \cite{HOPPET} 
and \cite{Pegasus} agree to five significant digits over a wide range in $x$ 
and $\mu_{\rm f}^2$, a level of agreement not reached before between 
\mbox{$x$-space} and $N$-space programs. 
The results at the important scale $\mu_{\rm f}^2 = 10^{\:\! 4}$ GeV$^2\,$ have
therefore been cast into reference tables for the evolution at LO, NLO -- 
including, for different initial conditions, the spin-dependent case -- and 
NNLO for the scales ratios $\mu_{\rm r}/\mu_{\rm f}^{} \, = 0.5,$ 1 and 2, 
using both a fixed and a variable number of flavours $\nf$ (see below). For 
example, the (iterated, see Ref.~\cite{Pegasus}) four-flavour NNLO evolution of
Eq.~(\ref{AVeq5}) for $\mu_{\rm f}^{} \, = \, 2 \mu_{\rm r}$ yields
$$
  x    \,=\, 10^{-5} \; , \quad
  xu_v \,=\, 2.9032 \cdot 10^{-3}\; , \quad \ldots \; , \quad
  xg   \,=\, 2.2307 \cdot 10^{\:\! 2} \qquad
$$

\vspace{-3mm}
\noindent\hspace*{1.9cm}$\ldots$

\vspace{-6mm}
\bea
  \;\; x \,=\, \; 0.9 \;\;\; , \quad
  xu_v   \,=\, 3.6527 \cdot 10^{-4}\; , \quad \ldots \; , \quad
  xg     \,=\, 1.2489 \cdot 10^{-6} 
\eea
at this scale. The complete tables can be found in Refs.~\cite{PDF-ref}. 
It would be very reasonable to employ only programs which have been checked 
against these benchmarks.

\section{Input shapes, factorization schemes and heavy quarks}
 
The \MSb\ scheme adopted so far is calculationally convenient and leads to a
perturbatively stable parton evolution -- recall Eq.~(\ref{AVeq4}), Figs.\
\ref{AVfig4} and \ref{AVfig7}. However, the NLO, \mbox{NNLO, \dots} parton 
distributions are not physical in this scheme. Therefore \MSb\ may not be the
scheme in which the initial distributions retain their physically motivated
shapes (as long known for the photon structure \cite{Gluck:1991ee}), e.g., for 
the proton's gluon density at large $x$, see Ref.~\cite{Martin:2004ir}.
Moreover, it seems unclear which positivity bound in particular $g(x,\mu^2)$
has to obey in this scheme, and which NLO partons (if any) are best suited for
obtaining estimates from leading-order Monte-Carlo programs \cite{NLO-MC}.

\vspace{1mm}
The traditional alternative to \MSb\ has been the DIS scheme 
\cite{Altarelli:1979ub}, in which the quark distributions are rendered physical
via the structure function $F_2$. 
For the singlet sector the transformation to this scheme is given by
\bea
\label{AVeq7}
  q_{S}^{\:\!\rm DIS} &\! =\! & q_{S}^{} \, +\, \as
     \big[ \, c_{\:\! 2,q}^{\:\! (1)} \otimes q_{S}
     \, +\, c_{\:\! 2,g}^{\:\! (1)} \otimes g \,\big] \, +\, \ldots
  \nn \\
  g^{\:\!\rm DIS}     &\! =\! & \; g\:\: -\, \as
     \big[ \, c_{\:\! 2,q}^{\:\! (1)} \otimes q_{S}
     \, +\, c_{\:\! 2,g}^{\:\! (1)} \otimes g \,\big] \, +\, \ldots
  \:\: .
\eea
Its large drawback is that the second row of Eq.~(\ref{AVeq7}) is arbitrary
except for the moment $N=2$ fixed by the momentum sum rule. Thus there is 
nothing physical about the \mbox{DIS-scheme} gluon density especially where 
constraints are needed most, for very large and for small $x$.

\vspace{1mm}
This shortcoming is absent in an interesting old suggestion, the DIS$_\phi$
scheme going back (at least) to Ref.~\cite{Furmanski:1981cw}. Here also the 
shape of the gluon distribution is rendered physical via the structure function
$F_\phi$ of a scalar directly coupling to gluons (such as the Higgs boson in 
the large-$m_{\rm top}$ effective theory). The transformation to DIS$_\phi$
at N$^{\rm n}$LO requires the corresponding coefficient functions 
$c_{\phi,q}^{\:\! (n)}$ and $c_{\phi,g}^{\:\! (n)}$. Scalar-exchange DIS
had to be considered anyway in Ref.~\cite{Vogt:2004mw}, and the determination 
of these coefficient function to order $\as^3$ requires only a minor extension
of the published calculations. These functions and possible constraints 
arising, for example, from the positivity of $F_\phi$ will be presented~%
elsewhere.

\vspace{1mm}
Now we turn to heavy quarks. 
For processes at a sufficiently high scale, charm and bottom become effectively
light flavours which have to be included in the parton structure of the proton.
For most values of $x$ one can disregard a possible non-perturbative `intrinsic
charm' (or bottom) component (which, however, can be relevant at large $x$ for
some specific LHC processes~\cite{CTEQ-H}). 
The \MSb\ evolution of $\as$ \cite{Chetyrkin:1997sg} and the parton densities 
with a variable number of flavours then proceed via a matching of effective 
theories for different $\nf$. The matching conditions for the parton 
distributions are especially simple at the heavy-quark mass, $\mu_{\rm f}^{} 
= m_h^{}$. Denoting the light-quark distributions by $l_i$, they up to 
N$^{m=2}$LO read \cite{Buza:1996wv}
\bea
\label{AVeq8}
 l_i^{\:(N_{\rm f}+1)} \quad &\! = \!\! &
   \, l_i^{\:(N_{\rm f})} \, +\, \delta_{m2} \:
   a_{\rm s}^2\: A^{\rm NS,(2)}_{qq,h} \otimes l_i^{\, (N_{\rm f})} 
 \nn \\[0.1mm]
 g^{\, (N_{\rm f}+1)} \quad &\! = \!\! &
   g^{\, (N_{\rm f})} \, +\, \delta_{m2} \, a_{\rm s}^2\, \big[
   A_{\rm gq,h}^{\rm S,(2)} \otimes q_{S}^{\, (N_{\rm f})} +
   A_{\rm gg,h}^{\rm S,(2)} \otimes g^{\, (N_{\rm f})} \big ] 
 \nn \\[0.1mm]
 (h+\bar{h})^{\, (N_{\rm f}+1)} \! &\! =\!\! & \hspace{1.35cm}
   \delta_{m2} \, a_{\rm s}^2\: \big [
   A_{\rm hq}^{\rm S,(2)} \otimes q_{S}^{\, (N_{\rm f})} +
   A_{\rm hg}^{\rm S,(2)} \otimes g^{\, (N_{\rm f})} \big  ] \:\: .
\eea
The results \cite{Buza:1995ie} underlying the $qq$, $hq$ and $hg$ 
coefficients have been confirmed recently \cite{Bierenbaum:2007qe}. 

\vspace{1mm}
The matching conditions (\ref{AVeq8}) are included in the above evolution codes
and benchmarks. Note that the $\as^2$ NNLO discontinuities were so 
far ignored in the (published) MRST parton densities. However, they have 
now been implemented and found to significantly affect the cross sections for 
$W/Z$ production at the LHC \cite{Martin:2007bv}. Forthcoming updates of also
the NLO distributions will include further significant improvements, e.g., the
use of fastNLO~\cite{Kluge:2006xs} for jet cross sections instead of 
pre-calculated $K$-factor tables.

\vspace{1mm}
In general, the calculation of heavy-quark effects on observables is far more 
involved.
We briefly summarize this issue for charm production at HERA, a process which
affects the vital extraction of the small-$x$ quark and gluon densities 
from $F_2^{\:\!p}$. There are three regimes: \linebreak
For $\,Q \gg \hspace*{-3.8mm} / \;\: m_c\,$ only $u$, $d$, $s$ and $g$ act as 
partons, and charm production can be calculated \linebreak 
using the fixed-order massive coefficient functions, presently known to 
NLO~\cite{F2CHARM}. This framework is usually referred to as the fixed-flavour 
number scheme (FFNS). 
At \mbox{$Q \gg\!\!\gg m_c$} all terms with $m_c/Q$ are negligible, and 
$\nf = 4$ partons -- obtained via the matching conditions~(\ref{AVeq8}) -- can 
be used with massless four-flavour coefficient functions, a procedure often 
called the zero-mass variable flavour-number scheme (ZM-VFNS). 
Finally there is, in general, an intermediate region $Q \gg m_c$, where terms
with $m_c/Q$ are not negligible, but large quasi-collinear logarithms require
a resummation via Eqs.~(\ref{AVeq2}). Then the $\nf = 4$ partons have to be 
used with `interpolating' coefficient functions for which several prescriptions
have been suggested, see refs.~\cite{Buza:1996wv,Chuvakin:1999nx,RT-CHARM,%
Tung:2006tb}. This is the genuine (or general-mass, GM-) VFNS.
 
\begin{figure}[bth]
\vspace*{-4.5mm}
\centerline{\epsfig{figure=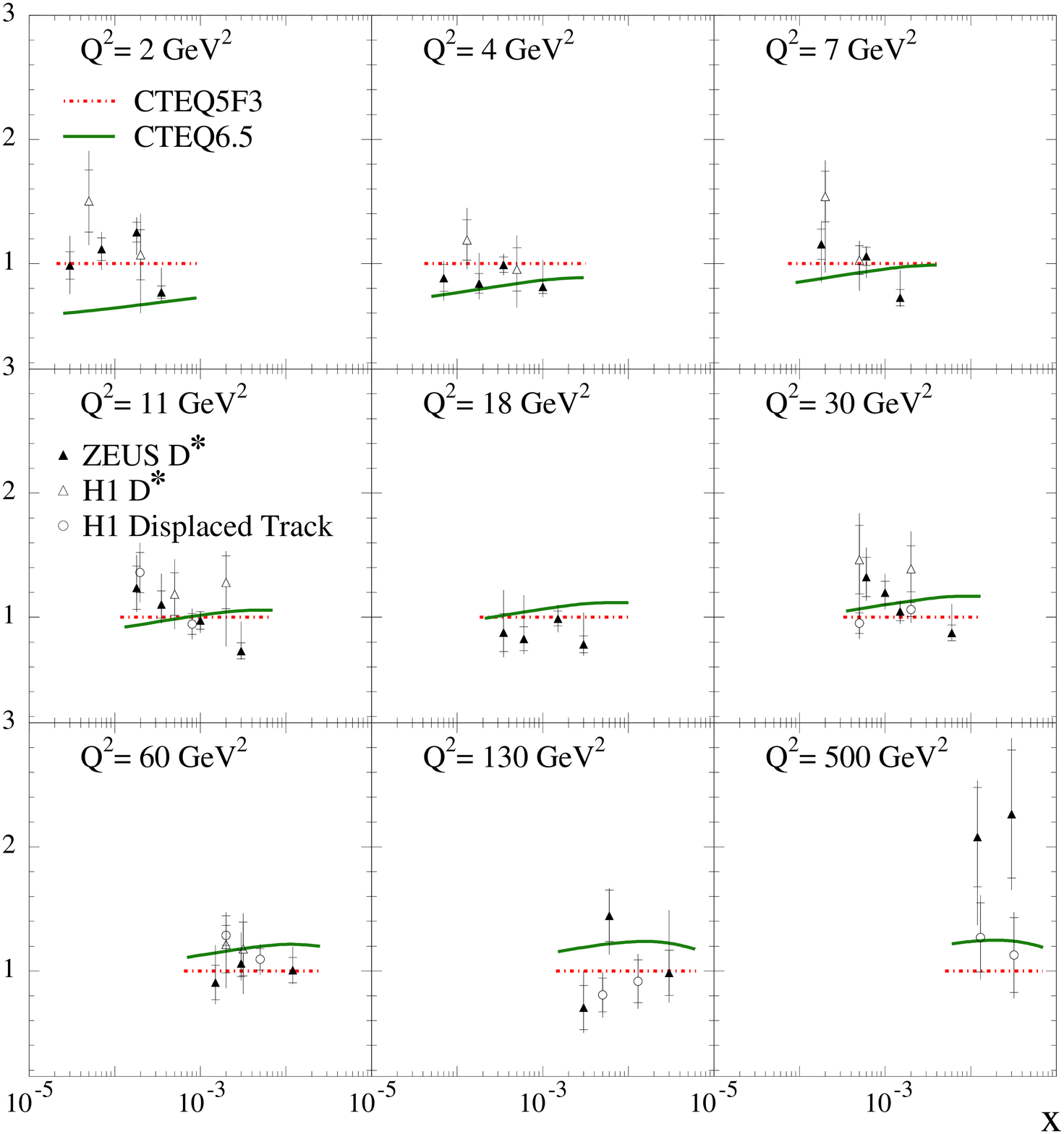,width=11.5cm,height=9.3cm}%
\hspace*{2mm}}
\vspace{-2mm}
\caption{HERA measurements of the charm structure function 
$F_2^{\:\!c_{}}$, compared to NLO CTEQ calculations in the fixed 
\cite{Lai:1999wy} and variable flavour-number \cite{Tung:2006tb} schemes. 
All results have been normalized to the former calculation (adapted from 
Ref.~\cite{Thompson:2007mx}).}\label{AVfig9}
\vspace*{-1.5mm}
\end{figure}

\vspace{1mm}
The transition regions between these regimes are process-dependent and tend to
lie at higher scales than one might at first expect, something to be kept in
mind when using bottom distributions with massless coefficient functions at the
LHC. 
For example, there are strong experimental (see Fig.~\ref{AVfig9}) and 
theoretical (recall, e.g., Ref.~\cite {Vogt:1996wr}) indications that the FFNS 
is applicable for the small-$x$ HERA data on $F_2^{\:\!c_{}}$ at least up to 
$\,Q^2 \gsim 100 \mbox{ GeV}^{\:\!2}$.

\section{Recent parton analyses and future LHC constraints}

\begin{figure}[bth]
\vspace*{0.5mm}
\centerline{\epsfig{figure=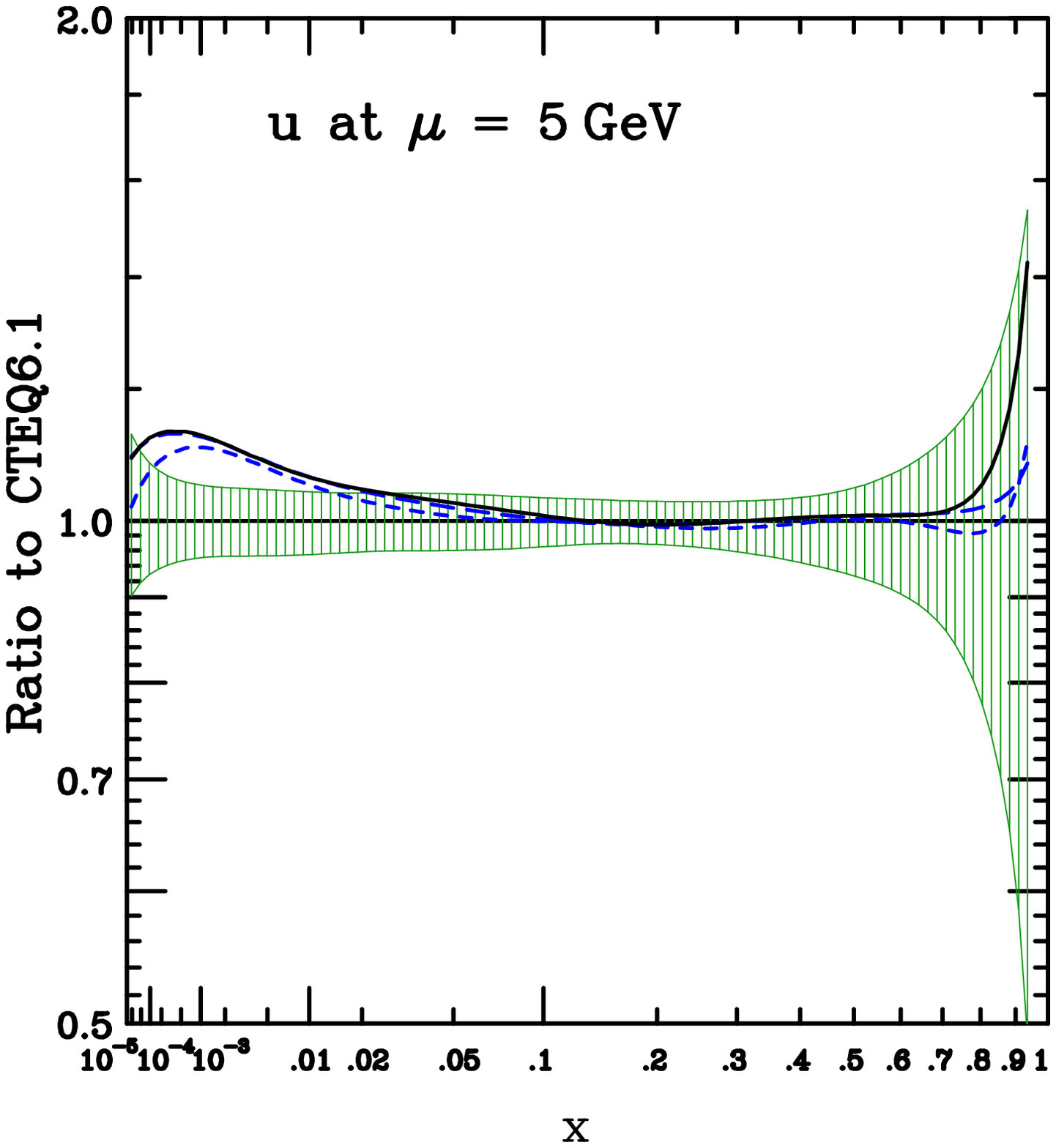,width=5.8cm}~~
\epsfig{figure=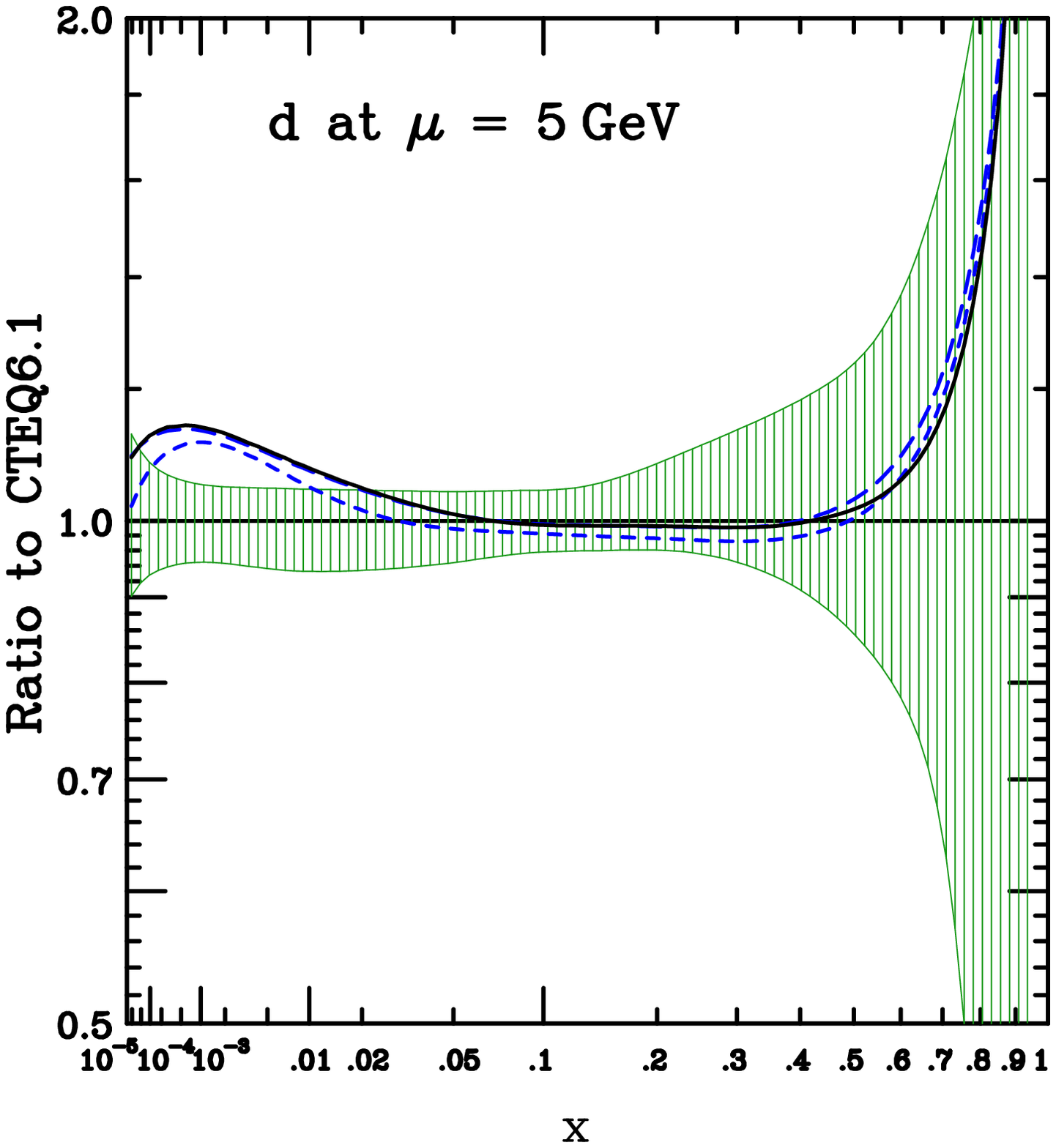,width=5.8cm}\hspace*{3mm}}
\vspace{-2mm}
\caption{Central $u$ and $d$ distributions of the CTEQ$\,$6.5 fit, normalized 
 to previous results from the same group \cite{Pumplin:2002vw}. 
 Also shown are the estimated error bands resulting from the experimental 
 uncertainties of the data included in the analysis (from Ref.~\cite
 {Tung:2006tb}).}\label{AVfig10}
\vspace*{-1mm}
\end{figure}

Recently the CTEQ collaboration has published a major update, CTEQ$\,$6.5, of 
their NLO global fits \cite{Tung:2006tb}. A salient improvement is that the 
mass suppression of the charm contribution to $F_2^{\:\! p}$ at HERA has 
finally been included -- before the inadequate ZM-VFNS (see above) had been 
used. The reduced charm component is compensated by larger $u$ and $d$ 
distributions at small $x$ as illustrated in Fig.~\ref{AVfig10}. This
increase leads to larger predictions for the $W$- and $Z$-production cross 
sections at the LHC, by about 8\%, a shift well outside the uncertainty bands 
obtained from the previous CTEQ$\,$6.1 sets \cite{Pumplin:2002vw}. It should be
noted, however, that both this shift and the similar NNLO result of Ref.~\cite
{Martin:2007bv} mentioned above do not invalidate the widths of the previous 
error bands. Rather the previous central values should be considered 
unreliable, as they resulted from fits disregarding well-known theoretical
information.

In any case, it is important to have at one's disposal several independent sets
of parton distributions at each order of perturbative QCD. Until recently, the 
only NNLO analysis besides those of MRST (now MSTW) was that of Ref.~\cite
{Alekhin:2002fv}, based only on data from deep-inelastic scattering. Last year 
this analysis has been expanded in Ref.~\cite{Alekhin:2006zm}: a consistent 
subset has been included of the available data on Drell-Yan lepton-pair 
production -- note the difference in approach to CTEQ, who are working on their
treatment of inconsistent data sets. The NNLO corrections to these cross 
sections \cite{Anastasiou:2003ds,Melnikov:2006di} are found to be crucial for 
the fits and, interestingly, as before a rather low value of $\as(M_Z)$ is 
preferred, in marked contrast to the recent NNLO fits of 
MSTW~\cite{Martin:2007bv}.

\vspace{1mm}
Usually the initial conditions for Eq.~(\ref{AVeq2}) are written in terms of an
ansatz, as in Eq.~(\ref{AVeq5}) but with more free parameters. The resulting
bias is monitored by varying this functional forms as, e.g., in the two dashed 
curves in Fig.~\ref{AVfig10}. An alternative approach is pursued by the NNPDF
collaboration, using neural networks to avoid any such bias. A first 
analysis of non-singlet structure functions has been performed in Ref.\
\cite{DelDebbio:2007ee}, using a new hybrid evolution method combining 
advantages of the $x$-space and Mellin-$N$ techniques mentioned above.

\begin{wrapfigure}{r}{0.55\columnwidth}
\vspace*{-3.5mm}
\centerline{\epsfig{figure=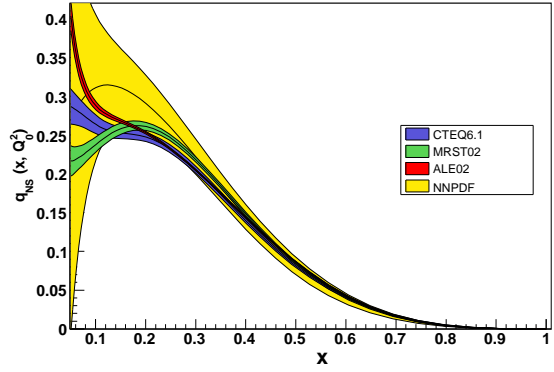,width=8cm,angle=0}%
\hspace*{-5mm}}
\vspace{-2.5mm}
\caption{Experimental error bands for the NLO non-singlet combination $\, 
q_{\;\!\rm ns}^{\;\! +} = u\!+\!\bar{u} - (d\!+\!\bar{d}\,)$, \mbox{according} 
to the older fits in Refs.~\cite{MRST0102,Pumplin:2002vw,Alekhin:2002fv} and 
the recent NNPDF analysis (from Ref.~\cite{DelDebbio:2007ee}).}\label{AVfig11}
\vspace*{-3mm}
\end{wrapfigure}

Four uncertainty bands for the combination $\,u + \bar{u} - (d + \bar{d}\,)
\,$ of NLO quark distributions are displayed in Fig.~\ref{AVfig11}. 
There are many differences between the chosen analyses of 
Refs.~\cite{MRST0102,Pumplin:2002vw,Alekhin:2002fv,DelDebbio:2007ee},
thus it seems difficult to isolate the possible impact of the parametrization
bias. It would be interesting to see fits of a reference data set using  
different approaches to the initial conditions but otherwise identical 
conditions.  In any case, given the precision of the data on the proton
structure function $F^{\:\!p}_2$ and the neutron-proton ratio, for example at 
$x \approx 0.2$, it seems rather unlikely that the very wide NNPDF 
band reflects the true uncertainty. 

\vspace{1mm}
Finally a non-singlet analysis of electromagnetic DIS has been performed in 
Ref.~\cite{Blumlein:2006be}, besides the quark distributions focusing on 
determinations of $\as$ up to the N$^3$LO of Eq.~\ref{AVeq3}. This order is accessible outside 
the small-$x$ region since, as confirmed by Ref.~\cite{Baikov:2006ai}, the
N$^3$LO corrections to the structure function evolution are dominated by the 
known coefficient functions, see Fig.~20 of Ref.~\cite{Vermaseren:2005qc}.
The results of Ref.~\cite{Blumlein:2006be} for the strong coupling constant 
read $\as(M_Z) = 0.1134, 41 \pm 0.002\,$ at $\,$N$^{2,3}$LO, consistent with 
Ref.~\cite{Alekhin:2006zm} but not with Ref.~\cite{Martin:2007bv}. 
\mbox{Obviously} more research is required before firm conclusions can be drawn 
on the uncertainties of the parton densities (as in Fig.~\ref{AVfig11}) and 
the determination of $\as$.

\begin{figure}[bht]
\vspace*{1.5mm}
\centerline{\psfig{figure=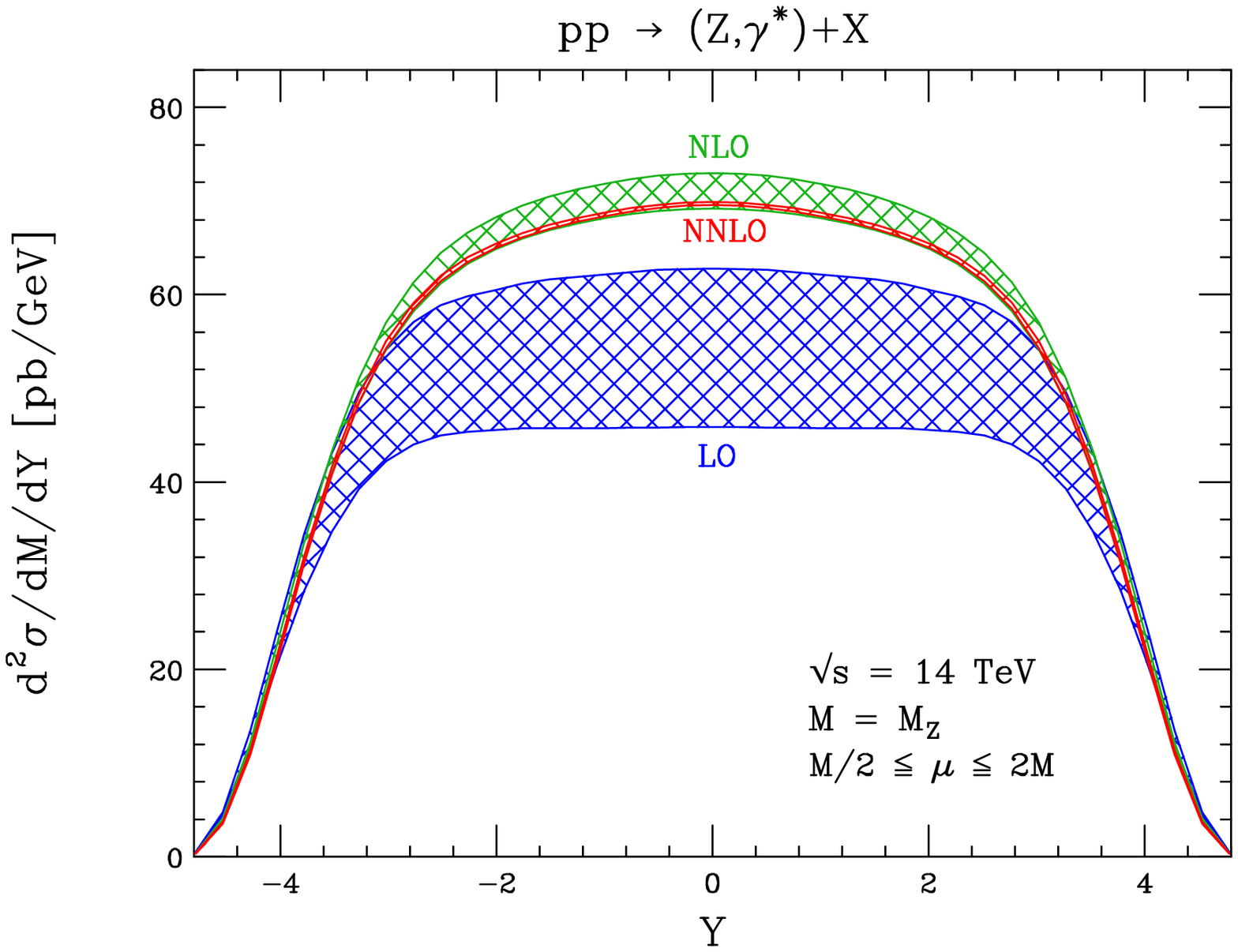,height=5.5cm,width=6.6cm}
\hspace*{4mm}%
\psfig{figure=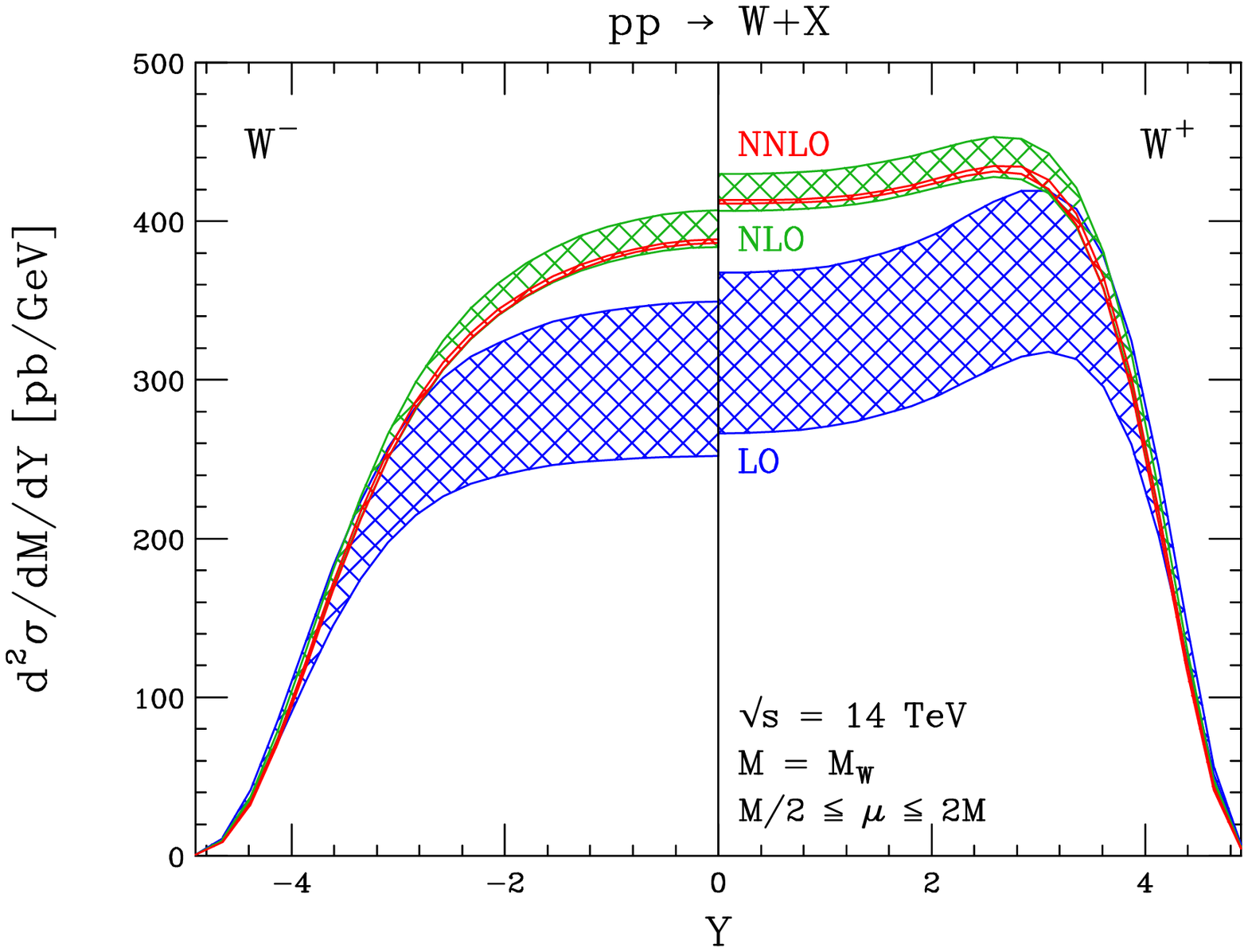,height=5.5cm,width=6.6cm}\hspace*{2mm}}
\vspace{-2mm}
\caption{Rapidity-dependent cross sections for gauge-boson production at the 
LHC, using the partons of Ref.~\cite{MRST0102}. Shown are the 
theoretical uncertainty estimates obtained by varying the scale $\mu$ by the
arbitrary but conventional factor of two around $M_{W,Z}$
(from Ref.~\cite{Anastasiou:2003ds}).}\label{AVfig12}
\vspace*{-0.5mm}
\end{figure}

\vspace{1mm}
The pre-LHC determinations of the parton densities will be improved upon by 
including reference cross sections measured at the LHC. The `gold-plated' 
process of gauge-boson production is illustrated in Fig.~\ref{AVfig12}; see 
Ref.~\cite{CooperSarkar:2005na} for a more detailed discussion including 
experimental aspects. The results shown demonstrate the importance of NNLO 
results even for processes with a far more benign perturbative expansion 
than the Higgs-production cross section of Fig.~\ref{AVfig2}: It would clearly 
be impossible to make precision predictions, or perform precision analyses,
based on the rough (and non-overlapping) LO and NLO error estimates obtained 
by varying the renormalization and factorization scale(s). Thanks to the NNLO
calculations \cite{Anastasiou:2003ds,Melnikov:2006di}, on the other hand, the 
perturbative uncertainty has  been reduced to a level of about 1\%, an accuracy
unprecedented for hadron-collider cross sections. 

\vspace{1mm}
\section{Outlook: HERA results for the LHC era}

Precision parton densities and QCD cross sections are required to fully realize
the potential of the LHC. For example, a very precise $W$-mass determination 
with \mbox{$\delta M_W \lsim 10$ MeV} seems experimentally feasible, see 
Ref.~\cite{LHC-MW}. Combined with $\delta m_{\rm top} \simeq 1$ GeV such a 
result could help to discriminate between, e.g., the standard model and its
minimal supersymmetric extension -- see the figure (updated from Refs.~\cite
{Heinemeyer:2006px}) shown at the end of the talk \cite{url}. While great 
progress has been made during the past years, considerable challenges remain.

\vspace{1mm}
At the time of this write-up 15 years of data-taking at HERA have ended.
Its results will remain indispensable throughout the LHC era, and it is 
important that also the high-luminosity results of the last phase are fully 
exploited -- despite the obvious temptation to move on to, say, LHC Higgs 
hunting as soon as possible. Moreover, it is highly desirable to preserve 
important data, e.g., on heavy quarks and jet production, in a manner 
facilitating detailed re-analyses (as performed for PETRA data in Ref.~\cite
{JADEnew}) a decade from now.

\vspace{1mm}
\section*{Acknowledgements}

It is a pleasure to thank M. Botje and S. Heinemeyer for providing figures
used here and in Ref.~\cite{url}, and S. Moch and R. Thorne for critically
reading parts of the manuscript.


\begin{footnotesize}

\end{footnotesize}


\end{document}